# A Man Misunderstood: Von Neumann did not claim that his entropy corresponds to the phenomenological thermodynamic entropy

Erin Sheridan

Department of Physics and Astronomy, University of Pittsburgh, Pittsburgh, PA 15213

**Abstract** Recently, attention has returned to the now-famous 1932 thought experiment in which John von Neumann establishes the form of the quantum mechanical von Neumann entropy $-\text{Tr}\rho\ln\rho$ ($S_{VN}$), supposedly by arguing for its correspondence with the phenomenological thermodynamic entropy ($S_{TD}$.) Hemmo and Shenker (2006) reconstruct von Neumann's thought experiment and argue that it fails to establish this desired correspondence. Prunkl (2019) and Chua (2019) challenge Hemmo and Shenker's result in turn. This paper aims to provide a new foundation for the current debate by revisiting the original text (von Neumann (1996, 2018)). A thorough exegesis of von Neumann's cyclical gas transformation is put forth, along with a reconstruction of two additional thought experiments from the text. This closer look reveals that von Neumann's goal is *not* to establish a link between $S_{VN}$ and $S_{TD}$, as is assumed throughout the current debate, but rather to establish a correspondence between $S_{VN}$ and the Gibbs statistical mechanical entropy $S_G$. On these grounds I argue that the existing literature misunderstands and misrepresents his goals. A revised understanding is required before the success of von Neumann's reversible gas transformation can be definitively granted or denied.

**Acknowledgements** I would like to thank Professors John Norton and David Wallace for recommending this topic of study, for their careful review of the text and for the helpful discussions.

**Contents**

It is my impression that von Neumann's monograph- which is today after more than eighty years, more frequently cited than ever before- is perhaps more frequently cited than actually read, and is my certain knowledge that it contains treasures and provides valuable insights that are too seldom appreciated.

– Nicholas Wheeler, *Preface to the New Edition of Mathematical Foundations of Quantum Mechanics* (2018)

# 1. INTRODUCTION

In his seminal work of 1932, *Mathematical Foundations of Quantum Mechanics*, John von Neumann rigorously investigates the foundational problems of quantum statistical mechanics and the nature of the measurement process. In **Chapter V**, von Neumann uses statistical and thermodynamic methods to better understand the irreversibility of the quantum mechanical measurement process. In a famous (and now disputed) thought experiment, he performs a cyclical, or reversible, transformation on an ideal quantum gas, and from the process' reversibility, derives the form of what we now know as the von Neumann entropy $S_{VN} = -\text{Tr}\rho \ln \rho$ by proving its correspondence to classical entropy.

This result is called into question by Hemmo & Shenker (H&S), who perform their own version of von Neumann's thought experiment in the one-particle, (multiple) finite-particle and infinite-particle cases (Hemmo & Shenker, 2006). H&S conclude that the von Neumann entropy successfully corresponds to the phenomenological thermodynamic entropy *only* in the thermodynamic limit, i.e., in a system with an infinite number of particles. Since infinite-particle systems do not exist in nature, von Neumann fails to establish the desired correspondence between the two entropies.

Chua (Chua, 2019) argues in response that the single-particle reconstruction employed by H&S is extraneous to the question of the correspondence of $S_{VN}$ and $S_{TD}$. While he agrees with H&S that von Neumann's goal is to establish the correspondence of the von Neumann entropy with the phenomenological thermodynamic entropy, he argues that since this is von Neumann's goal, the only instance of the thought experiment that matters is the infinite-particle case. H&S's single-particle construction is then, in addition to being an invalid thermodynamic system, completely irrelevant. Irrelevance aside, Chua maintains that H&S's argument against the correspondence of the two entropies in the single-particle case is *factually incorrect,* i.e., that H&S count entropies incorrectly.

I begin in **Section 2** by summarizing von Neumann's original quantum gas transformation thought experiment. I then take a closer look at two additional thought experiments in the original text (von Neumann, 1996, 2018), the first being the expansion of a single-particle gas, and the second being the macroscopic detection of photons on photographic plates. This closer look shows how von Neumann differentiates between two distinct classical entropies: the "classical entropy" $S_{CL}$ and the "macroscopic classical entropy," $S_M$. I argue that $S_{CL}$ is the entropy to which $S_{VN}$ is compared in the reversible gas transformation and the single particle gas expansion examples, whereas $S_M$ is treated separately in the photon detection example. I conclude my exegesis in **Section 2.4** by arguing that $S_{CL}$ corresponds to the Gibbs statistical mechanical entropy $S_G$, while $S_M$ is von Neumann's true analog to the phenomenological thermodynamic entropy $S_{TD}$. My main argument, based on these findings, is that von Neumann does not aim to identify $S_{VN}$ and $S_{TD}$ in his famous and now controversial thought experiment.

In **Section 3** I review certain aspects of this debate in the literature. I carefully review H&S's single-particle version of von Neumann's thought experiment and their conclusions. Afterwards I briefly summarize Chua's and Prunkl's (Prunkl, 2019) critiques of H&S, focusing on the single-particle case.

In **Section 4** I use the insights from **Section 2** to argue that H&S (4.1), Chua (4.2) and Prunkl (4.3) either misinterpret or misrepresent the true goal of von Neumann's cyclic quantum gas transformation. I argue that Prunkl comes closest to understanding von Neumann's true goal, while failing to apply this understanding to her argument. The debate over whether or not $S_{VN}$ corresponds to $S_{TD}$ is based on an incorrect understanding of von Neumann's goals, and must modify its course accordingly. I conclude in **Section 5**.

In order to provide as faithful an analysis of his work as possible, my exegesis will use the notation von Neumann himself used in 1932. While this unfamiliar notation is different from modern bra-ket and density operator notation, and may appear cumbersome, I make an effort to clearly explain the meaning behind the equations used. As there exists debate over the validity of von Neumann's principal result, it is of the utmost importance to follow his original work as closely as possible. We must minimize the opportunities for artificial additions and misinterpretations to find their way into our analysis.

## 2. REVISITING VON NEUMANN'S ORIGINAL TEXT

### 2.1: Von Neumann's thought experiment

What motivates von Neumann to dedicate an entire chapter to thermodynamic considerations in a book about the foundations of quantum mechanics? The answer is simple: measurement. When we consider the question "What exactly *is* quantum mechanical measurement, and what exactly does it *do* to the systems we measure?" we find fundamental answers from within the domain of thermodynamics and statistical mechanics. More specifically, von Neumann asks

*"What happens to a mixture with statistical operator U, if a quantity $\mathfrak{R}$ with the operator R is measured on it?"* (347)

at the start of **Chapter V** of his *Mathematical Foundations of Quantum Mechanics*. To answer this question, let us review how a quantum mechanical mixture is defined.

We define an operator $R$ for the quantity $\mathfrak{R}$ with a complete orthonormal set of eigenfunctions $\phi_1, \phi_2, \ldots$ and corresponding eigenvalues $\lambda_1, \lambda_2, \ldots$. The statistical operator $U$ (which corresponds to the modern density matrix operator) measures the quantity $\mathfrak{R}$ in each element of an ensemble. The sub-ensembles in which the operator $R$ has the values $\lambda_n$ are collected to obtain a mixture with the statistical operator $U'$ as defined below.

Von Neumann aims to understand why the process of quantum mechanical measurement on a mixture ("**Process 1**") is irreversible, while unitary time-evolution following the Schrödinger equation ("**Process 2**") is reversible. Let's formally define both processes.

**Process 1**: Measurement of an ensemble.

$$U \to U' = \sum_{n=1}^{\infty} (U\phi_n, \phi_n) P_{[\phi_n]} \tag{1}$$
$$= w_n P_{[\phi_n]}$$

Where $w_n = (U\phi_n, \phi_n)$ is the inner product of the statistical operator $U$ with the state $\phi_n$, i.e. $w_n$ is the probability of that state. After a measurement, the fraction $w_n$ of the original ensemble has the value $\lambda_n$ of operator $R$. $P_{[\phi_n]}$ is the properly normalized statistical operator for state $\phi_n$. Next, let us look at another kind of intervention on the original mixture.

**Process 2**: Unitary time-evolution by the Schrödinger equation.

$$U \to U_t = e^{\frac{-i}{\hbar}tH} U_0 e^{\frac{i}{\hbar}tH} \tag{2}$$

Where $H$ is the Hamiltonian operator, here assumed to be time-independent, and $U_t$ refers to a mixture of several states, for example $P_{[\phi_t^{(1)}]}, P_{[\phi_t^{(2)}]}, \ldots$ with weights $w_1, w_2, \ldots$ .

These two processes, or interventions, are fundamentally different from one another. Why? How is it that **Process 2** does not increase the statistical uncertainty of mixture $U$, while **Process 1** does? Why is **Process 1** able to transform states into mixtures, whereas **Process 2** only transforms states into states?

To answer these questions, von Neumann turns to thermodynamic considerations. In a now famous and, as of late, controversial thought experiment (depicted in **Figure 1**), he demonstrates the correspondence between the classical entropy $S_{CL}$ and the quantum mechanical entropy $S_{VN}$ of a statistical mixture. In this thought experiment, we will study a Gibbs ensemble of identical, noninteracting systems $[S_1, \ldots, S_N]$ with the statistical operator $U$. We refer to this ensemble as a $U$-ensemble.

The systems $S_1, \ldots, S_N$ comprise an Einstein gas, which is defined as follows: each system $S_i$ is confined to a box $K_i$ with impenetrable walls, and each of these $K_1, \ldots, K_N$ boxes constitutes a "molecule" in the overall Einstein gas, which itself is contained in a very large container **K**. The volume $V$ of **K**, and temperature of a heat reservoir $T$ in contact with **K**, are such that the $[S_1, \ldots, S_N]$- gas, or $U$-gas, can be treated as an ideal gas. When we want to measure the states of the systems $S_1, \ldots, S_N$ we will have to "open" the $K_1, \ldots, K_N$ boxes.

Von Neumann is very careful to convey the statistical mechanical nature of the Einstein gas in use here. He reiterates:

"*184: for the following, the statistical nature of these* [thermodynamic] *laws is of chief importance.*" (359)

and

*"In the terminology of classical statistical mechanics, we are dealing with a Gibbs ensemble; i.e., the application of statistics and thermodynamics will be made not on the (interacting) components of a single, very complicated mechanical system with many (only imperfectly known) degrees of freedom[185] — but on an ensemble of very many (identical) mechanical systems, each of which may have an arbitrarily large number of degrees of freedom, and each of which is entirely separated from the others, and does not interact with any of them.[186]*

*185: This is the Maxwell-Boltzmann method of statistical mechanics […]*

*186: This is the Gibbs method […]. Here the individual system [within each box $K_i$] is the entire gas, and many replicas of the same system (i.e., of the same gas) are considered simultaneously, and their properties are described statistically."* (360)

In other words, we are not considering an individual, complicated mechanical system which consists of many interacting molecules, as in Maxwell and Boltzmann's statistical gas theory. We are dealing instead with a Gibbs ensemble of many replicas of the same system. This detail is of the utmost importance as we move forward.

Now let's dig in to the experiment. We want to find the entropy excess of a U-gas (a mixture) with respect to the entropy of a $P_{[\phi_n]}$-gas (a pure state) under the same conditions. Remember, this is the entropy of a U-ensemble of $N$ individual systems.

We will move through this thought experiment step-by-step. To keep things simple we are using only two containers, but von Neumann generalizes this experiment to $n$ containers for $\phi_1, \phi_2, \ldots, \phi_n$ eigenstates. For additional guidance, the Appendix includes a description of each stage in von Neumann's original words.

**Stage 1:** To start, we have an ideal quantum gas of N molecules (as described above) at temperature $T$. The gas is composed of a mixture of $P_{[\phi_1]}, P_{[\phi_2]}, \ldots$ gases of $w_1 N, w_2 N, \ldots$ molecules, respectively and is in a container **K** with volume $V$.

**Stage 2**: We then add a second container **K'** next to **K**, with the same volume $V$. We are going to replace the impermeable, unmovable wall between **K'** and **K** with two walls. The first is a movable, completely impermeable wall and the second is a semi-permeable wall[1]. This semi-permeable wall is transparent for the molecules in state $\phi_1$ but opaque for molecules in all other states[2]. Finally, we add a second semi-permeable wall on the right side of container **K**. This second semi-permeable wall is opaque for the molecules in state $\phi_1$ but transparent for the molecules in all other states.

---

[1] The following operation, according to von Neumann, is performed at the site of the semi-permeable wall: "We construct many windows in the wall, each of which is defined as follows: each "molecule" $K_1, \ldots, K_N$ of our gas (we are again considering U-gases at the temperature $T > 0$) is detained there, opened, the quantity $\Re$ measured on the system $S_1$ or $S_2$ or $\ldots S_N$ contained in it. Then the box is closed again, and according to whether the measured value of $\Re$ is $< 0$ or $> 0$, the box, together with its contents, penetrates the window or is reflected, with unchanged momentum." (368)

[2] Before beginning the thought experiment (pages 368-369), von Neumann proves that if the eigenstates $\phi_1, \phi_2, \ldots, \phi_n$ form an orthonormal set, then there is a semi-permeable wall which lets system $S_i$ pass through unhindered while reflecting each of the other orthogonal states unchanged.

**Stage 3**: Next, the movable, completely impermeable wall and the $\phi_1$-impermeable wall are pushed to the left, such that the distance between them is kept constant until the impermeable wall hits the left side of container **K'**. This ensures that no work is being done against the pressure of the gas.

**Stage 4**: The movable walls are then replaced by rigid, completely impermeable walls, so that **K'** and **K** are completely separated. At this stage, all $\phi_1$ molecules have been transferred from **K** into **K'** without any work being done and without any change in temperature.

**Stage 5**: Now we isothermally compress the outer walls of **K'** and **K** to volumes $w_1 V, w_2 V$, respectively, so that the gases in either container have the same density ($N/V$) and the total volume returns to the original volume $V$. To accomplish the isothermal compression, we must apply the amounts of work $w_1 NkT \ln w_1, w_2 NkT \ln w_2, \dots$ to the gases, and transfer the same amount of energy, as heat, to the heat reservoir. The entropy increase for the gas for this process is $\Delta S = \sum_{n=1}^{\infty} w_n Nk \ln w_n$.

**Stage 6**: Finally, we transform each of the $P_{[\phi_1]}, P_{[\phi_2]}, \dots$ gases into a of $P_{[\phi]}$-gas, where $\phi$ is an arbitrarily chosen state. Von Neumann is careful to establish the reversibility of this transformation, which is equivalent to a unitary reset of an $\mathfrak{R}$-measuring device. He refers to an earlier section (Section 2) in **Chapter V**, in which he takes great care to prove that a measurement transformation between two different U-ensembles (such as the $P_{[\phi_1]}, P_{[\phi_2]}, \dots$ gases and $P_{[\phi]}$-gas here) is reversible, i.e. "accomplished without the absorption or liberation of heat energy" (364). The key to achieving this reversibility is to remember that we are working with a Gibbs statistical ensemble many replicas of the same system, instead of a single system, and to make many measurements on the U-ensemble:

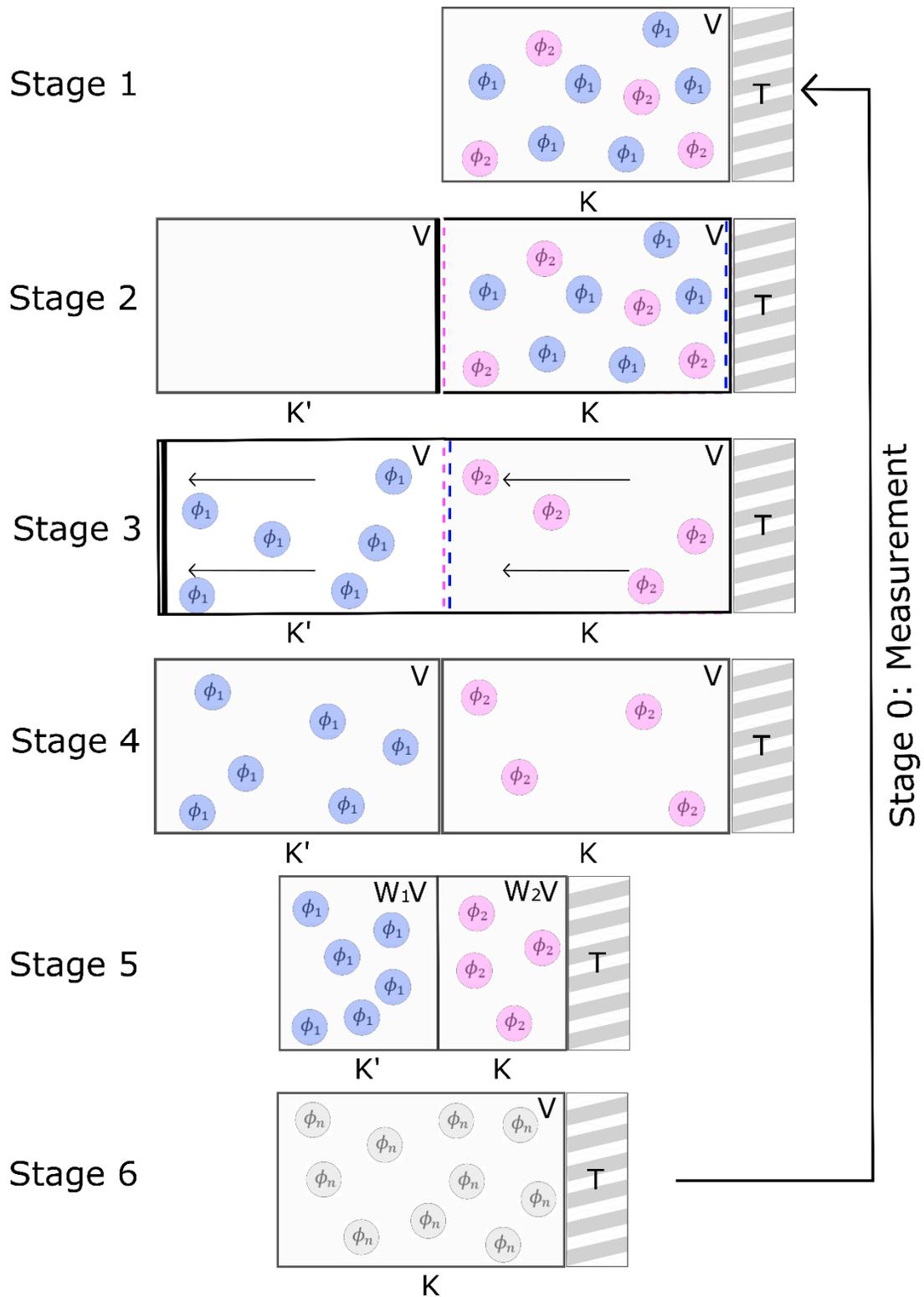

**Figure 1**. Von Neumann's original thought experiment: the reversible transformation of an ideal quantum gas. I have divided the reversible process into seven stages, Stage 0-Stage 6.

"*By repetition of a great number of different measurements we shall change $P_{[\phi]}$ into an ensemble which differs from $P_{[\psi]}$ by an arbitrarily small amount.*" (365)

Von Neumann proceeds to prove this mathematically, by taking the limit as number $k$ of measurements goes to infinity $k \to \infty$.

Now that the reversibility of the transformation is established, the wall dividing **K** and **K'** is (also reversibly) removed, and the $P_{[\phi]}$-gases are mixed, which is possible since the gases are identical and at equal densities. We end up with a $P_{[\phi]}$-gas of N molecules in a volume $V$. It is at this stage that von Neumann concludes the thought experiment, claiming that the desired reversible process is complete. Since the entropy increased by $\sum_{n=1}^{\infty} w_n Nk \ln w_n$ in **Stage 5**, and $S = 0$ in the final stage, the entropy must have been $-\sum_{n=1}^{\infty} w_n Nk \ln w_n$ in **Stage 1**. We can easily see that $\text{Tr}(U\ln U) = \sum_{n=1}^{\infty} w_n \ln w_n$, so that the entropy of a U-ensemble is $-Nk\text{Tr}(U\ln U)$.

**Stage 0**: For the sake of clarification, I have added **Stage 0** into von Neumann's thought experiment. In **Stage 0**, we perform a measurement (**Process 1**) on the $P_{[\phi]}$-gas that we have at the end of **Stage 6**. This measurement has the effect of transforming our pure state into a mixed state, the state we have in **Stage 1**, thereby increasing the entropy of the gas from $S = 0$ in **Stage 6** to $S = -\sum_{n=1}^{\infty} w_n Nk \ln w_n$ in **Stage 1**.

We have now completed a reversible process. Since the state of the quantum gas is identical in **Stage 1** at the start and the end of the cycle, the entropy change for the cycle must be $\Delta S = 0$. The only change in classical entropy $S_{CL} = \sum_{n=1}^{\infty} w_n Nk \ln w_n$ comes from the isothermal compression performed in **Stage 5**. The classical entropy does not change during **Stage 3**, because no work is done against the gas pressure when the impenetrable and semi-permeable walls are pushed to the left with a constant distance kept between them. The entropy of the ideal quantum gas must then be $S_{VN} = -\sum_{n=1}^{\infty} w_n Nk \ln w_n = -Nk\text{Tr}(U\ln U)$ for the mixed state in **Stage 1**, so that $\Delta S = S_{CL} + S_{VN} = 0$. We may then conclude that quantum mechanical measurement is dissipative by the amount $\Delta S_{VN}$.

## 2.2: Expansion of a single-molecule gas

Next, let us consider both the reversible and irreversible expansion of a single-particle ideal gas in a container, in contact with a heat reservoir at temperature $T$, in the style of Szilard's single-particle engine, (Szilard, 1929), as von Neumann does in Section 4 of **Chapter V** (**Figure 2**).

What is the purpose of this example? Recall that von Neumann's motivation for finding the entropy $S_{VN}$ of a quantum mechanical mixture was to explain why **Process 1**, the measurement of a mixture, is irreversible, while **Process 2**, time-evolution of the mixture under Schrödinger dynamics, is reversible. Though, as we've seen, von Neumann goes to great trouble to show that $S_{VN}$ corresponds to the classical entropy in the cyclic gas transformation, the original problem is still not solved. So we press on:

*"The situation is best clarified if we set [the number of molecules] $M = 1$. Thermodynamics is still valid for such a one-molecule gas, and it is true that its entropy increases by $k \ln 2$ if its volume is doubled."* (399)

We must consider two possible initial situations for the one-molecule ideal gas expansion. In **Initial Situation 1**, we do not know if the molecule is on the left or the right side of the container to start. We only know that it is initially found in a volume $V/2$. If we remove the partition separating the left (L) and right (R) compartments, the gas diffuses into the other side of the container. Its volume increases from $V/2$ to $V$, and the corresponding thermodynamic entropy change is $\Delta S = k \ln \frac{V}{V/2} = k \ln 2$, with no corresponding entropy change in the heat reservoir. The process is therefore irreversible.[3]

---

[3] N.B. Von Neumann's understanding of the expansion of a single-particle gas, as detailed here, differs from how single particle gas expansion is commonly understood in contemporary physics discussions. For example, if we do not know the location of the single particle to start (Initial Situation 1) the single particle has a 50% chance of being on the left side of the partition before it is removed, it also has a 50% probability of being on the left side after it is removed. The removal

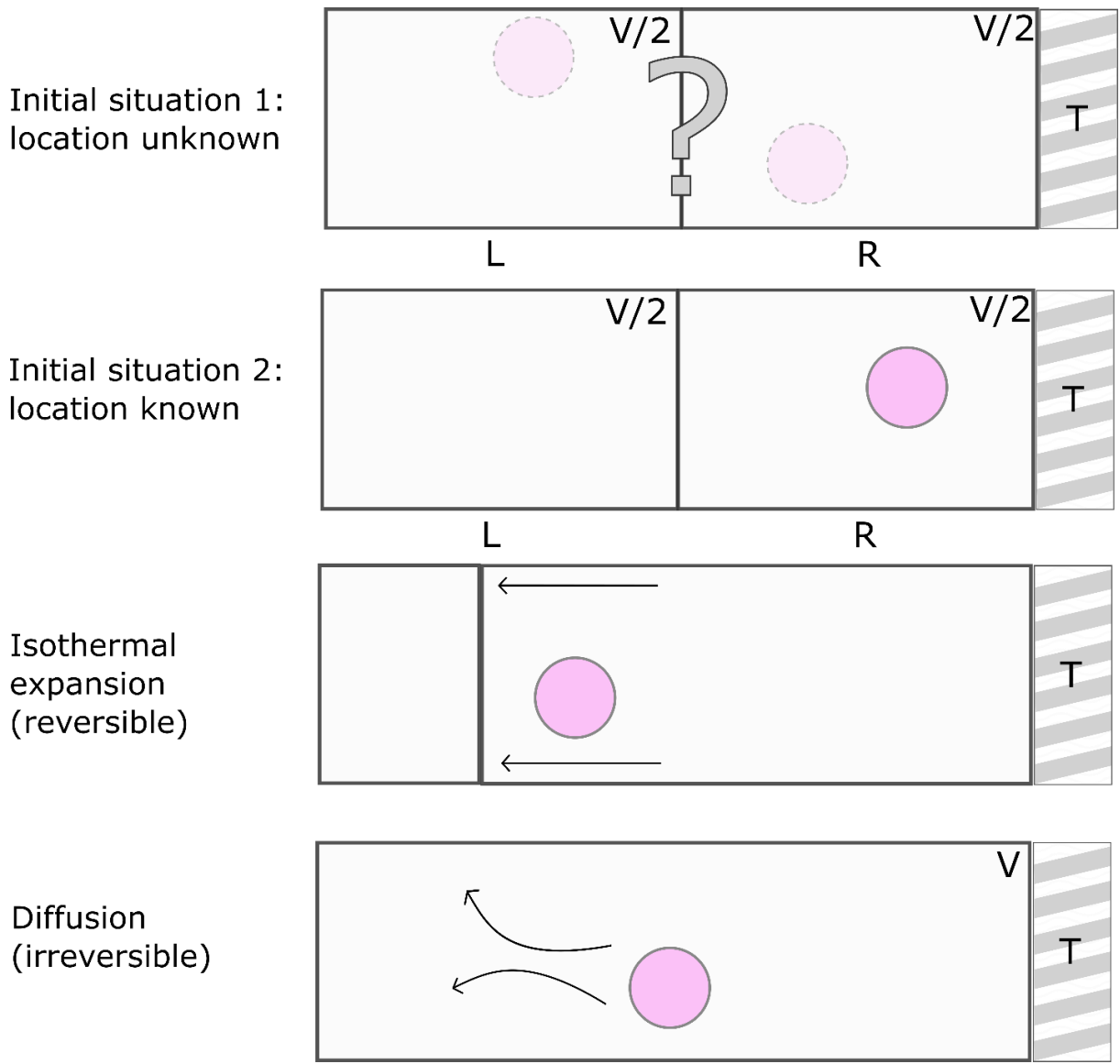

**Figure 2**. Von Neumann's description of the reversible and irreversible expansion of a single-molecule gas.

---

of the partition causes no entropy change, and the process is reversible. On the other hand, when we *do* know on which side the particle is location to start (Initial Situation 2), the particle's probability distribution changes, and the process is irreversible. Indeed, these different understandings are the subject of some controversy. Since our goal is exegetical, we will not address this issue, but we will move ahead assuming von Neumann's understanding: the process starting from Initial Situation 1 is irreversible, and the process starting from Initial Situation 2 is reversible. Thank you to David Wallace and John Norton for their instructive comments on this matter.

In **Initial Situation 2**, on the other hand, we know that the one-molecule gas is on the R side of the container. Since we know the molecule's initial location, the process of removing the partition and letting the gas diffuse is equivalent to allowing the gas to expand isothermally, and reversibly, by pushing against the partition until the gas volume increases from $V/2$ to $V$. The entropy change of the gas in this case is $\Delta S = k \ln 2$.

This process requires work; the corresponding entropy change for the heat reservoir is $\Delta S = -k \ln 2$. The overall entropy change for the system is zero and the process is reversible. Notice that we achieve the same final state as if we started in **Initial State 1**, i.e., at the end of the expansion, we no longer know whether the molecule is on the left or the right side of the container, but here, the overall entropy change is zero. As von Neumann notes, in this case

*"we have exchanged our knowledge for the entropy decrease $k \ln 2$."* (400)

In other words, the entropy of the system is the same for a single-molecule gas in either volume $V/2$ or $V$, provided that we know, as in **Initial Situation 2**, in which half of the container the molecule is found.

In classical physics, measurements do not increase a system's entropy. Instead, the entropy increases as a rule during ordinary mechanical evolution of the system. It is this apparent paradox which von Neumann wants to resolve with his single-molecule gas example. He writes (bolding mine):

*"Although our entropy expression is, as we saw, **completely analogous** to classical entropy, it is still surprising that it is invariant under normal temporal evolution of the system (**Process 2**), and increases only in consequence of measurements (**Process 1**)."* (398)

The resolution of von Neumann's classical/quantum measurement paradox lies within the knowledge of the observer. The entropy change for the expansion process of a single-molecule gas, and therefore its reversibility, depends upon the observer's knowledge of a particular observable (here: location.) What about other observable quantities?

If a classical observer knows the position and momentum of the single molecule before diffusion, the location of the molecule can be found at any later time. The entropy then remains constant throughout the diffusion process. From this reasoning, von Neumann concludes that time-variations in entropy are a result of an observer's incomplete knowledge: If an observer were to know the positions and momenta of every particle in a system, the entropy of the system would remain constant over time, as it does under the quantum-mechanical **Process 2**. But since macroscopic observers may only observe macroscopic quantities, leaving many "measurables" unmeasured, classical entropy increases over time (bolding mine):

*"**The time variations of entropy are based then on the fact that the observer does not know everything**- that he cannot find out (measure) everything that is measurable in principle. His senses allow him to perceive only the so-called macroscopic quantities. But this clarification of the apparent contradiction mentioned at the outset **imposes upon us an obligation to investigate the precise analog of classical macroscopic entropy for quantum mechanical ensembles** [...]"* (401)

The single molecule gas expansion example firmly establishes von Neumann's "subjective" view of the classical entropy. In this example, as in the cyclic quantum gas transformation, the entropy of a system increases when our ignorance about the state of the system increases. But von Neumann realizes that this solution leads us to another challenge: what, then, is the classical analog of the quantum mechanical entropy for *macroscopic* systems? The third and final example, discussed below, provides the answer.

### 2.3: Photon detection with photographic plates

How does one classify the entropy of a system for an observer who can only measure macroscopic quantities, perhaps only with limited accuracy? According to von Neumann, all limited-accuracy measurements of a particular quantity (say, position) can be replaced with absolutely accurate measurements of different quantities which are functions of the original quantities. These new quantities allow us to calculate the classical macroscopic entropy $S_M$.

As a demonstrative example, von Neumann investigates the macroscopic measurement of two non-simultaneously measurable quantities: the position $q$ and momentum $p$ of two photons. These two quantities are not simultaneously measurable in quantum mechanics, but can be measured simultaneously if we limit the precision of our measurement. This limitation requires that $q$ is measured with light wavelengths that are greater than some minimum length, and $p$ is measured with light trains that are shorter than some maximum length.

The macroscopic measurement of photon position and momentum involves detecting two photons with photographic plates. The position $q$ is measured on one photon scattered via the Compton effect, and the momentum $p$ is measured on another photon that is first reflected, then has its frequency Doppler-shifted, and is finally deflected using a diffraction grating. At the end of the experiment, each photon will have produced a black spot on a separate photographic plate.

The macroscopic observer then measures the location of the black spot, or the spot-coordinates, left on each plate by each photon, with arbitrary precision. The spot-coordinates of the two black spots are simultaneously measurable, and are related to $q$ and $p$ with ineliminable error values $\epsilon$ and $\eta$, respectively, where $\epsilon\eta \sim h$. If we define operators $Q$ and $P$ for quantities $q$ and $p$, then we can also define macroscopic operators $Q'$ and $P'$ for the macroscopically measurable quantities $q'$ and $p'$ for the spot coordinates of either black spot.

Let us define the eigenvalues of the precisely measurable macroscopic operator $P'$ as $\lambda^{(n)}$. A macroscopic observer may ask the question, "Is $P' = \lambda^{(1)}$?" These questions become the observer-specific macroscopic projections $E_n$. The $E_n$ correspond to macroscopically answerable questions whose quantities $\mathfrak{E}$ have values 0 or 1. For example, we define the projection $E_1 = [\text{"Is } P' = \lambda^{(1)}?\text{"}]$ with the quantity

$$\mathfrak{E}(\lambda) = \begin{cases} 1, & \lambda = \lambda^{(1)} \\ 0, & otherwise \end{cases} \quad (3)$$

A set of projections $E_n$ completely characterizes a macroscopic observer and can be understood as a coarse-graining mechanism:

*"It should be observed that the $E_n$-which are the building blocks of the macroscopic description of the world- correspond in a certain sense to the cell division of phase space in classical theory."* (409)

Now we are ready to ask the question: what entropy does the mixture $U$ have for a macroscopic observer characterized by projections $E_1, E_2, ...$? If we define $s_n = \text{Tr}(E_n) \geq 1$, and the expectation value of $E_n$ for ensemble $U$ is $\text{Tr}(UE_n)$, then the macroscopic entropy of $U$ becomes

$$-k \sum_{n=1}^{\infty} \text{Tr}(UE_n) \ln\left(\frac{\text{Tr}(UE_n)}{s_n}\right). \quad (4)$$

The macroscopic entropy always changes with time[4], unlike the von Neumann entropy which is constant under unitary time evolution via **Process 2**, and it is never less than the von Neumann entropy of the mixture[5]:

---

[4] Von Neumann argues that the macroscopic entropy always changes with time because the Hamiltonian $H$ can never be a macroscopic quantity (since all macroscopic quantities must commute.) In other words, the energy can never be measured macroscopically with complete description. The interested reader is directed to von Neuman's footnote 204 on page 410.

[5] Recall that we defined $s_n = \text{Tr}(E_n) \geq 1$. We know this because "if all of the $s_n = 1$ [...] This would mean that macroscopic measurements would themselves make possible a complete determination of the state of the observed system. Since this is ordinarily not the case, we have in general $s_n > 1$ and in fact $s_n \gg 1$." (265)

$$-k \sum_{n=1}^{\infty} \mathrm{Tr}(UE_n) \ln\left(\frac{\mathrm{Tr}(UE_n)}{s_n}\right) \quad (5)$$

$$\geq -k\mathrm{Tr}(U\ln U)$$

With equality only when

$$U = \sum_{n=1}^{\infty} \frac{\mathrm{Tr}(UE_n)}{s_n} E_n. \quad (6)$$

Von Neumann attempts to prove this equality under certain conditions at the end of **Chapter V**. Whether or not this proof is successful is beside the point; the essential takeaway is that it is *here* where von Neumann argues for the equality of the von Neumann entropy and the macroscopic classical entropy (under particular conditions), not in the cyclic gas transformation thought experiment.

One question remains: does von Neumann argue here for the strict correspondence of the macroscopic classical entropy with the phenomenological thermodynamic entropy? No; he defers to Boltzmann's kinetic theory of gases here, noting that this theory requires additional statistical assumptions:

*"Since the macroscopic entropy is always time variable, the next question to be answered is this: **Does it behave like the phenomenological thermodynamics of the real world:** i.e., does is predominantly increase? The question is answered affirmatively in classical mechanical theory by the so-called Boltzmann H-theorem. In that, however, certain statistical assumptions- namely, the so-called 'disorder assumptions'-must be made."* (415)

## 2.4: Von Neumann's true goal

Recent debate in the literature centers around the question, "does the von Neumann entropy correspond to the phenomenological thermodynamic entropy?" with the implicit question, "is von Neumann's thought experiment successful?" Based on our investigations above, I argue that the first question is misguided and should not be used to answer the second; von Neumann does not aim to identify $S_{VN}$ with $S_{TD}$ in his thought experiment. Rather, he intends to prove the correspondence between $S_{VN}$ and the "classical entropy" $S_{CL}$, and $S_{CL}$ is most accurately interpreted as $S_G$. We see this is the case for three reasons.

First, most glaringly: we saw in **Section 2.1** that von Neumann unambiguously establishes the statistical mechanical nature of the quantum gas used in his cyclical gas transformation. The gas is not a single system but a finite statistical ensemble containing many replicas of the same system. It would seem, then, that even a single-particle Einstein gas consists of many multiple particles. We will revisit this fact in our discussion of Prunkl's paper in **Section 3.5**.

According to von Neumann, the reversible gas transformation proves that $S_{VN}$ is *completely analogous* to $S_{CL}$. By classical entropy, it appears that he does *not* mean phenomenological thermodynamic entropy $S_{TD}$. But how can that be the case, seeing that he uses macroscopic thermodynamic quantities, such as volume and temperature, in the thought experiment?

We answer with our second reason: in the first two thought experiments we studied, von Neumann follows Szilard's example and *assumes* the validity of thermodynamic laws in order to draw conclusions about the statistical mechanical entropy (or classical entropy) $S_{CL}$. He explicitly states this assumption early on in **Chapter V** (see also the quote included in **Section 2.1**):

*"First, let us assume the validity of both of the fundamental laws of thermodynamics, i.e., the impossibility of perpetual motion of the first and second kinds (energy law and entropy law), and proceed on this basis to calculation of the entropy of ensembles [...] the correctness of both laws will be assumed and not proved."* (359)

Von Neumann assumes the validity and applicability of thermodynamic laws in his examples, including the single-molecule gas expansion example, as Szilard does in his famous paper on Maxwell's demon. This assumption is not to be ignored.

When von Neumann writes that the single-molecule gas is a valid thermodynamic system (**Section 2.2**), he is not claiming that the single particle constitutes what we would refer to as a phenomenological thermodynamic system. In contemporary vocabulary, one would be justified in reading von Neumann's use of "thermodynamic" as "statistical mechanical" or "statistical thermodynamic" here.

Instead of asking "does $S_{VN}$ correspond to $S_{TD}$?" the sceptic is better off asking "is von Neumann justified in his use of macroscopic thermodynamic quantities in his cyclic gas transformation and single-particle gas expansion examples?" One could reasonably argue that von Neumann is not justified in doing so, just as one can argue that Szilard's use of thermodynamic quantities in his own thought experiment is not justified (see, for example, (Norton, 2016)).

Finally, we arrive at the third reason why von Neumann does not aim to identify $S_{VN}$ with $S_{TD}$ in the cyclic gas transformation. After resolving the measurement paradox with the single molecule gas expansion example (see **Section 2.3**), von Neumann tasks himself with defining the quantum mechanical analog of the *classical macroscopic entropy $S_M$*. He does this in the photographic plates example. The classical *macroscopic* entropy he defines here is clearly distinct from the classical entropy $S_{CL}$: $S_M$, *not* $S_{CL}$, is von Neumann's analog to the phenomenological thermodynamic entropy.

In the final paragraph of **Chapter V**, von Neumann finally asks the question we've been waiting for: does $S_M$ correspond to $S_{TD}$? Instead of answering this question himself, von Neumann defers to Boltzmann's -theorem and lets the answer rest on the validity of Boltzmann' statistical assumptions. In other words, the jury is out. Nowhere in his work does von Neumann argue for the direct correspondence of $S_{VN}$ and $S_{TD}$. $S_{CL}$ is not analogous to the phenomenological thermodynamic entropy, but rather to the "subjectivist" Gibbs statistical mechanical entropy $S_G$, which is apparent based on von Neumann's subjectivist interpretation of entropy in the single molecule gas expansion example.

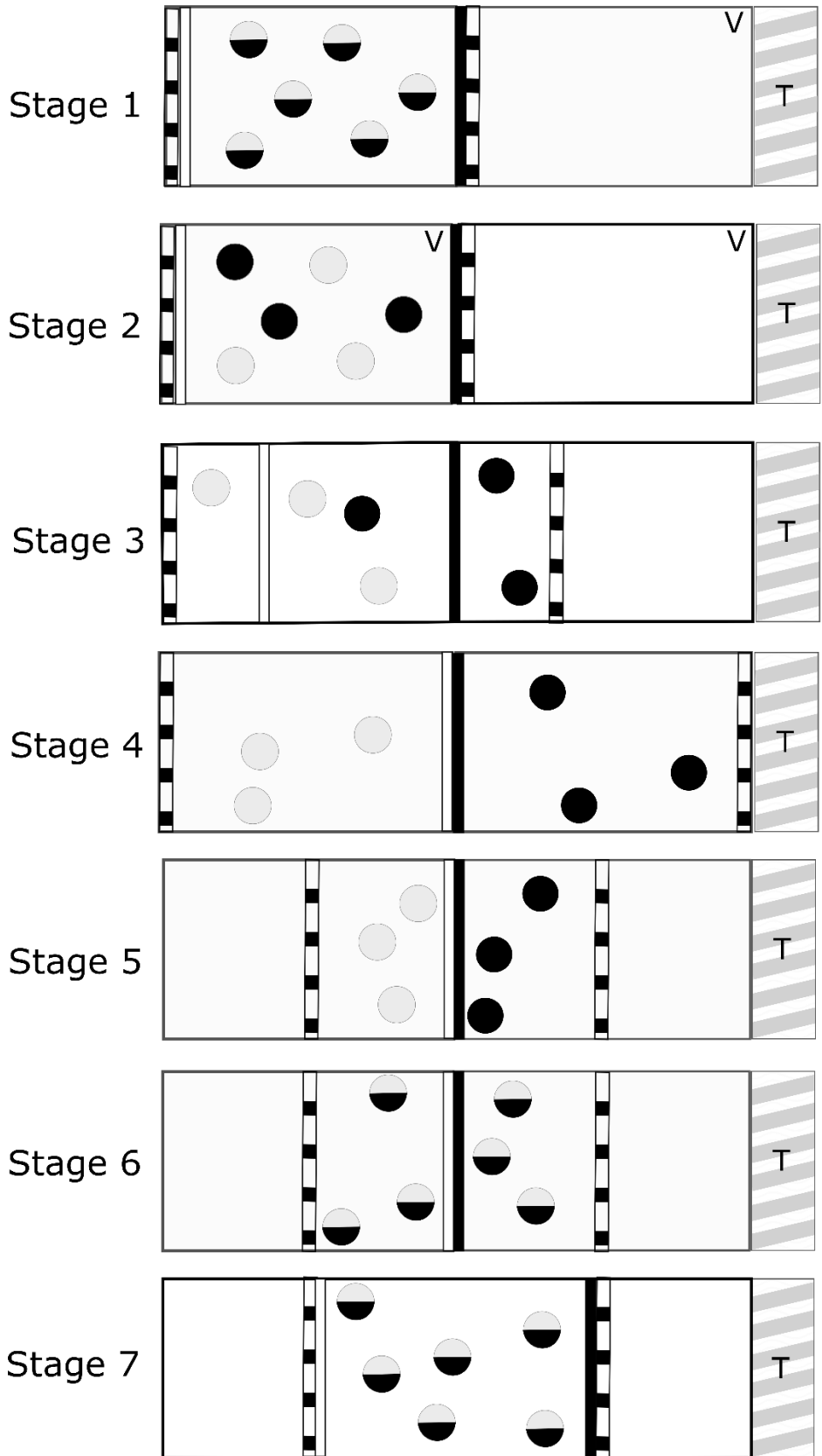

**Figure 3**. H&S's reconstruction of von Neumann's thought experiment.

Equipped with this reasoning, we will now survey the debate in the literature concerning the correspondence of $S_{VN}$ and to $S_{TD}$ (**Section 3**) and show that it is for the most part gravely misguided (**Section 4**). With this as our aim, we now review H&S's reconstruction and critique of von Neumann's thought experiment.

## 3. RECEPTION OF VON NEUMANN'S THOUGHT EXPERIMENT

### 3.1: Hemmo and Shenker's thought experiment

H&S begin their critique of von Neumann's reversible ideal quantum gas thought experiment with their own summary of it (see **Figure 3**), which proceeds as follows:

In place of the generalized operator $R$ with the orthonormal set of eigenfunctions $\phi_1, \phi_2,\ldots$ and corresponding eigenvalues $\lambda_1, \lambda_2, \ldots$, used by von Neumann, H&S's version of the experiment uses the two-level quantum mechanical operator for the z-spin degree of freedom. **Stage 2** of their experiment, equivalent to **Stage 1** of von Neumann's, is prepared by performing a z-spin measurement on a collection of spin-1/2 particles. The particles are initially prepared in the up eigenstate of the x-spin operator, so that in **Stage 2** we have a quantum mechanical mixture of particles in either the up or down eigenstate of the z-spin operator.

The experiment proceeds as described in **Section 2.1** above, with a slightly different labeling of stages. Importantly, H&S make sure to point out that in **Stage 5**, when isothermal compression of the quantum gas occurs, the pressure in both compartments of the container must become equal. In order for this to happen, a measurement of the relative number of particles in each compartment of the container is necessary. (Remember that in **Stage 6** in **Section 2.1**, we required the density of the quantum gas in all compartments to be the same: $N/V$.) H&S argue that von Neumann fails to mention this second, required measurement; this omission ultimately leads to the failure of his thought experiment.

## 3.2: The single-particle case

H&S spend the majority of their paper discussing their single-particle version of von Neumann's thought experiment (**Figure 3**). Their ultimate conclusion is that von Neumann's argument fails to "establish the conceptual linkage between $\text{Tr}\rho \ln \rho$ and the thermodynamic quantity $\frac{1}{T}\int pdV$ (or $dQ/T$) in the case of a single particle gas." We'll briefly outline the steps for this process, according to H&S, below.

**Stage 1**: A particle $P$ is prepared in the spin-up eigenstate of the x-spin operator, in the left container. The quantum state for the particle is written as

$$\rho^{(1)} = |+_x\rangle\langle+_x|_P \rho(L)_P |\text{Ready}\rangle\langle\text{Ready}|_M$$

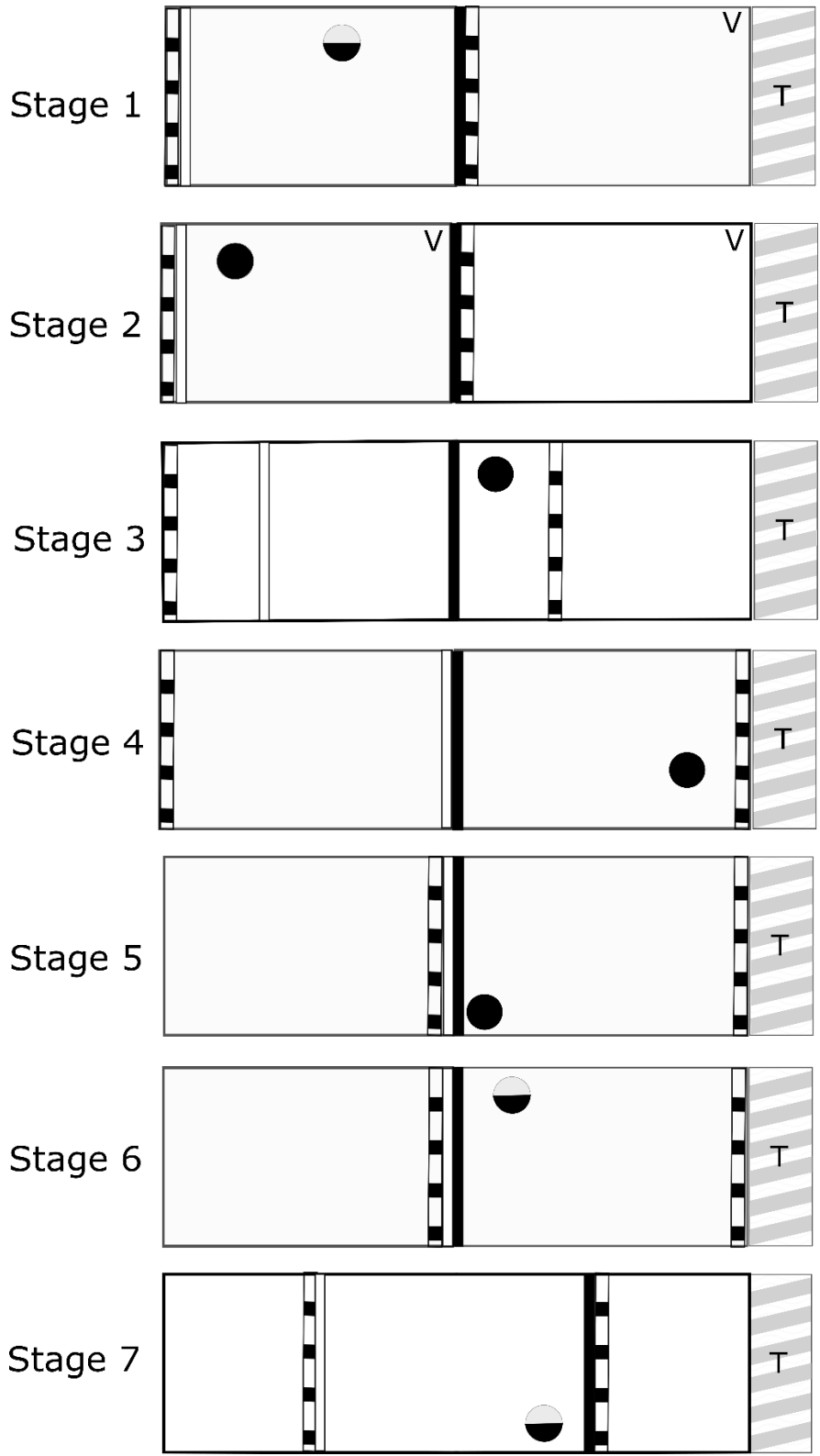

**Figure 4.** H&S's single-particle reconstruction.

Where $\rho(L)_P$ is the density matrix for the position of $P$, the coarse-grained position is either Left (L) or Right (R), and $P$ is measured by the measuring instrument $M$, which begins in the Ready state. Here $S_{VN} = 0$.

**Stage 2**: The z-spin of $P$ is measured, and the quantum state becomes

$$\rho^{(2)} = \frac{1}{\sqrt{2}}(|+_z\rangle\langle+_z|_P \rho(L)_P |+\rangle\langle+|_M + |-_z\rangle\langle-_z|_P \rho(L)_P |-\rangle\langle-|_M)$$

After the z-spin measurement, the z-spin of the particle becomes entangled with the state of $M$. However, H&S trace out the measuring device, obtaining a reduced quantum state:

$$\rho^{(2,red)} = \frac{1}{2}(|+_z\rangle\langle+_z|_P + |-_z\rangle\langle-_z|_P)\rho(L)_P$$

and summarize: "This state has the *form* of a classical mixture, which in some interpretations of quantum mechanics may be taken to describe our ignorance of the z-spin of $P$." It is important to clarify here that this state is not a quantum mechanical superposition, but a mixed state. Here $S_{VN} > 0$.

**Stage 3**: As described above, impermeable and semi-permeable walls are inserted and moved, keeping their separation constant, in order to move the molecule in the up-eigenstate to the left and the down-eigenstate to the right.

**Stage 4**: After spatial separation is complete, the reduced quantum state of $P$ is

$$\rho^{(4,red)} = \frac{1}{2}(|+_z\rangle\langle+_z|_P \rho(L)_P + |-_z\rangle\langle-_z|_P \rho(R)_P)$$

Indicating that the z-spin of the particle is now correlated with its position.

**Stage 5**: As before, we perform isothermal compression so that the total volume of the system returns to the original volume $V$, while the pressure becomes equal on both sides of the container. To achieve this, we must fully compress the empty side of the container to volume zero, while leaving the side of the container with the particle uncompressed.

Note that the empty side of the container is compressed against vacuum, so no work is required for the compression and the thermodynamic entropy of the system does *not* change, contrary to **Stage 5** of von Neumann's original thought experiment. This presents us with a major problem. If the thermodynamic entropy does not change in this stage, then the overall entropy change for the entire cycle will be nonzero, contradicting von Neumann's result.

It gets even worse. In order to carry out this isothermal compression, H&S maintain that we must know in which side of the container particle $P$ is located. We are therefore required to take a measurement of the particle's location prior to compression, and this measurement process adds new terms into the entropy arithmetic. H&S devote much attention to this location measurement, which we will revisit momentarily.

**Stage 6**: The particle is "reset" to the up eigenstate of the x-spin operator. The measurement device must also be returned to its initial Ready state, which can be done unitarily.

**Stage 7**: The partition separating the two sides of the container is lifted, returning the system to its original state. Stage 7 can also be carried out unitarily.

H&S's single-particle experiment raises some apparent issues with von Neumann's reasoning, particularly concerning the isothermal compression performed in **Stage 5**. A measurement of the particle's location is required prior to compression in this stage; a measurement von Neumann appears to ignore. And because compression will be performed against a vacuum, the thermodynamic entropy does not change in **Stage 5**. Von Neumann's demonstration of the equivalence between $S_{VN}$ and thermodynamic entropy consequently fails:

"*Therefore, whatever changes occur in $\text{Tr}\rho \ln \rho$ during the experiment, they cannot be taken to compensate for $1/T \int pdV$, since the latter is null throughout the experiment. For this reason, Von Neumann's argument does not establish a direct* *connection between and the thermodynamic entropy (in both the collapse and no-collapse theories) in the case of a single particle."* (162)

In the wake of this disturbing conclusion, we are left with two questions to consider. First, what is the physical significance of the location measurement performed in **Stage 5**? Second, does von Neumann really intend to establish a direct correspondence between $S_{VN}$ and *thermodynamic* entropy $S_{TD}$?

We start with the first question. H&S claim that the location measurement in **Stage 5**, while it has no effect on thermodynamic entropy, has an effect on $S_{VN}$. This effect differs in collapse and no-collapse interpretations of quantum mechanics. For collapse interpretations, the quantum state collapses to a pure z-spin state

$$\rho^{(4)} = |+_z\rangle\langle+_z|_P \rho(L)_P \text{ or } \rho^{(4)} = |-_z\rangle\langle-_z|_P \rho(R)_P$$

while in no-collapse theories, the quantum state remains a mixed z-spin state

$$\rho_4 = \frac{1}{2}\left(|+_z\rangle\langle+_z|_P \rho(L)_P + |-_z\rangle\langle-_z|_P \rho(R)_P\right).$$

Now, while this difference does not change the value of $S_{VN}$ or $S_{TD}$, it does change the value of what H&S refer to as the "information theoretic entropy" (which I label $S_I$):

*"In subjectivist approaches to statistical mechanics, entropy quantifies our knowledge regarding the gas, using e.g. the Shannon information."* (163)

H&S point out that, despite their consideration of $S_I$ as a means of correcting von Neumann's arithmetic, $S_I$ is *not* the quantity von Neumann considers in his original thought experiment. According to their argument, von Neumann clearly attempts to demonstrate the correspondence between $S_{VN}$ and $S_{TD}$, not between $S_{VN}$ and $S_I$. Nevertheless, since the inclusion of $S_I$ may redeem the thought experiment, H&S generously lay this issue aside and press onward.

During the location measurement in **Stage 5**, the Shannon information $\sum_i p_i \ln p_i$ increases by ln2, i.e. the information theoretic entropy $S_I$ decreases by $-\ln 2$. How does this additional entropy change affect the entropy arithmetic?

According to H&S, in collapse interpretations of quantum mechanics, both $S_{VN}$ and $S_I$ change during the location measurement in **Stage 5**. Our arithmetic depends on whether or not these two entropies are conceptually distinct. Recall that because the measurement returns a mixed state to a pure state here, $S_{VN}$ decreases by ln2 to $S_{VN} = 0$. Also recall that in **Stage 2**, $S_{VN}$ increases from zero to ln2. If we can identify $S_{VN}$ with $S_I$, the entropy increase in **Stage 2** is compensated exactly by the entropy decrease $-\ln 2$ in **Stage 5**. If $S_{VN}$ *cannot* be identified with $S_I$, however, we must take both of these changes into account and sum them, so that the overall entropy change in **Stage 5** is $-2\ln 2$. This change does not compensate for the change in **Stage 2**, and so von Neumann's argument fails.

To summarize, in collapse interpretations of quantum mechanics, von Neumann's argument holds as long as we assume $S_{VN} = S_I$ *and* $S_I = S_{TD}$. But, H&S argue, this amounts to assuming what the thought experiment sets out to prove. Since $S_{VN}$ cannot be identified with $S_I$, the thought experiment fails.

Finally, let us briefly turn our attention to no-collapse interpretations. As we saw above, in a no-collapse scenario, the location measurement in **Stage 5** does not change the reduced quantum state, so there is no change in $S_{VN}$. $S_I$ still increases upon measurement, because

*"relative to each component of the mixture, the observer acquires information, and this acquisition is expressed by a change in the Shannon information"* (164)

and this change in Shannon information exactly compensates for the increase in $S_{VN}$ in **Stage 2**. Von Neumann's argument then works out in no-collapse interpretations of quantum mechanics… sort of. While the arithmetic here succeeds in identifying $S_{VN}$ with $S_I$, this success falls short of von Neumann's goal of identifying $S_{VN}$ with $S_{TD}$, since a linkage between $S_I$ and $S_{TD}$ cannot be assumed.

We have been led to the second question above: does von Neumann really and truly set out to identify $S_{VN}$ with $S_{TD}$ in his original thought experiment? While I have argued that he does not, H&S argue that he does:

*"von Neumann's motivation in this argument was to establish a direct linkage between* $\text{Tr}\rho\ln\rho$ *and thermodynamic entropy."* (161)

I will address this disagreement in detail in **Section 4.1**.

### 3.3: The finitely- and infinitely-many particle cases

After considering the single-particle case, H&S discuss the finite-particle and infinite-particle cases of the reversible quantum gas transformation. Recall that for the single-particle case, a location measurement is required in **Stage 5** to determine the compartment in which the particle is located. When there are finitely many particles $N > 1$ in the container, we don't need to measure the location of each particle individually in order to perform quasi-static compression in **Stage 5**, but we still need to know the relative populations of particles in the left and right compartments. Unlike the single particle location measurement, in this case the relative populations can be measured in such a way that $S_{TD}$ remains unchanged.

When the left and right compartments are isothermally compressed to return to the original volume $V$, the change in $S_{TD}$ due to the compression depends on the relative distribution of the particles. We assume an equal right-left distribution or particles for large enough $N$, because it is highly probable. $S_{VN}$ corresponds to $S_{TD}$ only when the particles are equally distributed between the two compartments, because an equal left-right distribution negates the need of a location measurement. The entropy changes in **Stage 5** are then the same in both collapse- and no-collapse scenarios. Unfortunately, our assumption of equidistribution will never be exact for a finite number of particles:

"*Strictly speaking, no matter how large N may be, as long as it is finite, the net change of entropy throughout the experiment will not be exactly zero. Since* *equidistribution is not the only possible distribution, in principle one cannot dispense with the location measurement before the compression."* (169)

Von Neumann's argument fails here as well. So much for the finite-particle case.

H&S treat the infinite-particle case last. We can understand it straightforwardly based on the reasoning above. Put simply, we can assume an exact equidistribution of particles across the left and right compartments of the container at the exact infinite limit. Then von Neumann's argument goes through.

But not so fast! We must proceed with caution in the presence of infinities. H&S outline two ways to analyze this case (names mine):

(1) *Statistical method*: Here we take "a time series of identical experiments, such that as time goes to infinity the laws of large numbers imply that the relative frequencies of the outcomes of the experiments approach the theoretical probabilities […] we measure individual quantities of each of the particles separately and only then count the relative frequencies" (170)
(2) *Simultaneous method*: Here "all measurements are completed and the outcomes are all present at a given time […] we measure relative frequencies directly." (170)

The *Statistical method* considers the physical state of the gas as it approaches infinity, while the *Simultaneous method* considers the physical state of the gas at the infinite limit. H&S maintain that, because von Neumann's goal is to establish the correspondence between $S_{VN}$ and $S_{TD}$, it is crucial to his argument that we follow the *Simultaneous method*.[6]

---

[6] It is interesting that the *Statistical method*, which appears to be a faithful understanding of the nature of the Einstein gas used in von Neumann's original thought experiment, is ignored by H&S, while the *Simultaneous method*, which has no grounding in von Neumann's original text, is deemed crucial.

The system considered using this method is unphysical, since infinite systems do not exist. Regardless, H&S press on. After some work we arrive at the intuitive conclusion:

"at *the infinite limit there is no need to measure before the compression the relative frequencies in order to know* with certainty *that they are equal to the quantum mechanical probabilities (unlike the finite case.) Therefore, arithmetically von Neumann's argument goes through at the infinite limit*." (172)

It isn't enough to save von Neumann's argument. This case is not physical, and hence not permissible, because all real systems have a finite number of particles. In summary,

*"We saw that von Neumann's argument goes through only at the limit of infinitely many particles […] [it] does not hold for a very large or even enormous number of particles, and not as the number of particles approaches infinity, but only at the limit of an infinite number of particles. However, real systems are finite. This means that von Neumann's argument does not establish a conceptual identity between the Von Neumann entropy and thermodynamic entropy of physical systems."* (172)

## 3.4: Chua's argument against H&S

Like H&S, Chua addresses the issue of the correspondence of the von Neumann entropy and the thermodynamic entropy. He argues that H&S's critique against von Neumann fails. I break down his argument into three components, as follows:

(a) *Historical component*: Chua claims, contrary to H&S, that von Neumann's goal was not to establish a correspondence between $S_{VN}$ and $S_{TD}$ in all domains; rather, von Neumann's goal was to establish only an approximate correspondence between the two entropies that goes through in the thermodynamic limit. H&S's single- and finite-particle analyses are therefore irrelevant, and the fact that the correspondence between $S_{VN}$ and $S_{TD}$ fails for finitely many particles does not threaten the success of von Neumann's argument. $S_{TD}$ is the phenomenological thermodynamic entropy, after all, and thermodynamics requires the infinite-particle limit, so it is no shock that the argument might only succeed in this case. H&S do not see this because they suffer from "a confusion between phenomenological and statistical thermodynamics." (34)

(b) *Validity component*: Citing work by Norton (Norton, 2017) Chua argues that the single-particle case is thermodynamically unsound, and therefore extraneous to the question of the correspondence of $S_{VN}$ and $S_{TD}$:

"*Fluctuations relative to single-particle systems are large, and generally prevent these systems from being in equilibrium states at any point of the process, rendering reversible processes impossible in the single particle case.*" (17)

(c) *Arithmetical component*: Irrelevance aside, Chua argues that H&S's entropy calculation in the single-particle case is incorrect due to a misunderstanding of the nature of quantum mixed states. This confusion leads to an incorrect account of the entropy change associated with the location measurement in **Stage 5** of the single-particle thought experiment.

In **Section 4** I focus on the historical and validity components of Chua's argument, which center on the inadmissibility of the single- and finite-particle cases of the reversible quantum gas transformation. It is not our aim to determine the veracity of the arithmetical component here.

### 3.5: Prunkl's argument against H&S

Prunkl attacks H&S's argument from another angle. Unlike Chua, she does not explicitly deny the validity of the single-particle case as a representation of von Neumann's original thought experiment (i.e. the validity and historical components). Instead, Prunkl's main critiques center around the role of the measurement apparatus. I break down her argument into two components below, though it appears that her strongest argument against H&S's single-particle reconstruction is contained within a one-paragraph appendix on the statistical nature of von Neumann's quantum gas, also discussed below.

(a) *Landauer's Principle component*: Landauer's Principle states that there is a heat cost associated with resetting a measurement apparatus. Appealing to this principle, Prunkl challenges H&S's assumption that the measurement apparatus can be reset to its initial state unitarily in **Stage 6** of the single-particle experiment. The assumption of a unitary reset, she argues, leads to a violation of the Second Law of Thermodynamics. To be clear, the measurement device to be reset is the device which measures the z-spin of the particle and is reset to the up eigenstate of the x-spin operator in **Stage 6**.

(b) *Joint entropy component*: Notwithstanding this discussion, Prunkl finds another issue with H&S's single-particle experiment. She claims, somewhat along the lines of the *Arithmetical component* of Chua's argument, that H&S calculate $S_{VN}$ incorrectly. Specifically, H&S fail to consider the entropy contribution of the additional measurement apparatus required for the additional location measurement in **Stage 5**.[7] As a result, their single-particle experiment fails even when the Landauer's Principle critique is ignored.

Prunkl argues that the system's so-called conditional entropy decreases during the location measurement, while the joint entropy of the system combined with the additional measurement apparatus remains unchanged. When one remembers to take the measurement apparatus into account, using the joint entropy and not the conditional entropy (like H&S) in their entropy arithmetic,

"*the analogous behavior of thermodynamic entropy and von Neumann entropy for the joint system is restored.*" (8)

(c) *Statistical mechanical component*: Apart from the critiques summarized briefly above, and perhaps most importantly, Prunkl briefly discusses to the statistical

---

[7] Prunkl's argument in fact goes deeper than this brief description. Prunkl argues that H&S's Stage 5 location measurement is redundant, which H&S fail to see because, according to Prunkl, they misunderstand the nature of quantum mechanical measurement. After successfully arguing this point, Prunkl grants H&S this additional location measurement for the sake of argument.

mechanical nature of the Einstein gas used by von Neumann in the reversible gas transformation. In an Appendix, she writes:

*"In his original setup, von Neumann introduced a 'gas' consisting of individual systems, locked up in boxes and placed in a further, giant box. The 'gas' represents a imaginary* (sic) *statistical but finite ensemble[…] This means that even in the case of an individual quantum system, von Neumann's argument would remain unchanged: the density operator of this individual quantum system would still relate to an ensemble of systems and a system containing a single particle would therefore still be modeled as an N particle ensemble. The statistical representation of a) a system containing a single particle, and b) a system containing many particles, are therefore identical."* (20)

After this, Prunkl refers to von Neumann's single-molecule gas expansion example. In agreement with our exegesis above, she concludes, contrary to Chua,

*"This, however, does not imply that von Neumann denies the meaningful application of thermodynamics to individual particles, quite the contrary: von Neumann explicitly considers the case of a single particle in a box."* (20)

Prunkl has shown a deep insight into von Neumann's original work here. Unfortunately, this matter is left to the Appendix, and is not developed into an argument against H&S's reconsruction, as will be discussed more in **Section 4.3**.

## 4. VON NEUMANN'S GOAL IS MISUNDERSTOOD

### 4.1: H&S misunderstand von Neumann's goal

H&S consider the role of the "subjectivist entropy," or the information entropy, $S_I$ in their discussion of the single-particle thought experiment, despite their claim that von Neumann does not argue for the correspondence of $S_I$ and $S_{VN}$. In fact, H&S explicitly grant that $S_I$ could be considered relevant here:

"*the subjectivist approach may seem relevant to von Neumann's own discussion of his thought experiment since von Neumann himself mentions Szilard's (1929) paper on Maxwell's demon*" (163)

Given this excerpt, it is surprising that H&S fail to grasp the true goal of von Neumann's thought experiment. They remain firm in their view that

*"von Neumann's motivation in this argument was to establish a direct linkage between* $\text{Tr}\rho\ln\rho$ *and thermodynamic entropy."* (161)

We have demonstrated (**Section 2.4**), on the basis of (a) the statistical mechanical nature of the Einstein gas used in the reversible gas transformation, (b) the stated unproven assumption of the validity of thermodynamic laws in a statistical setting and (c) the definition of the classical macroscopic entropy $S_M$ in the photographic plates example, that von Neumann's true goal in the cyclic gas transformation is to establish a correspondence between $S_{VN}$ and the classical entropy $S_{CL}$, understood in subjectivist terms, i.e. as the Gibbs statistical mechanical entropy $S_G$.

After acknowledging von Neumann's single molecule gas example, why do H&S maintain their incorrect interpretation of von Neumann's goal? They seem to misunderstand its purpose. As shown in **Section 2.2**, von Neumann uses this example to resolve his measurement paradox. Under a subjectivist understanding of statistical mechanical entropy, he concludes from this example that **Process 1** increases entropy because the observer has incomplete knowledge of a system.

To the contrary, H&S maintain that von Neumann must argue for the correspondence of $S_{VN}$ and $S_{TD}$ in the cyclic gas transformation in order to resolve the measurement paradox:

*"recall von Neumann's original motivation in proposing this thought experiment. Von Neumann wanted to make use of the thermodynamic arrow of time in order to explain the irreversible behavior of the quantum state in a measurement (von Neumann's process 1)."* (165)

This interpretation has no basis in von Neumann's original work. As we've shown, von Neumann resolves the measurement paradox in a separate example, without having to appeal to phenomenological thermodynamics.

Furthermore, our exegesis of both von Neumann's and H&S's cyclic gas transformations reveals that the two experiments are not identical. The most glaring difference is the additional measurement step added in before the isothermal compression in **Stage 5**. So, not only must we call H&S's interpretation of von Neumann's goal into question; we must also question the faithfulness (and hence, the validity) of their reconstruction, and of **Stage 5** in particular.

## 4.2: Chua also misunderstands von Neumann's goal

On what does Chua base the historical component of his critique of H&S? He claims that von Neumann intended for $S_{VN}$ to work merely as an approximation of the phenomenological thermodynamic entropy:

*"Von Neumann's strategy was never to demonstrate the strict identity of $S_{VN}$ and $S_{TD}$, i.e. the correspondence of $S_{VN}$ and $S_{TD}$ in all domains. Instead, it was to show that between $S_{VN}$ corresponds to between $S_{TD}$ only in the domain where phenomenological thermodynamics hold, in all other cases merely approximating $S_{TD}$"* (32)

But nowhere in his work does von Neumann describe the von Neumann entropy as an approximation of any other kind of entropy. Indeed, after the completion of his reversible ideal quantum gas thought experiment, he writes that $S_{VN}$ is *completely analogous* to the classical entropy, not merely approximately analogous, nor analogous only in the thermodynamic limit. As I argued in **Section 2.3** what von Neumann means by "classical entropy" is the statistical mechanical entropy. This becomes apparent when von Neumann works through the expansion of a single-molecule ideal gas. As a result, we cannot dismiss H&S's single-particle and finite-particle cases as irrelevant, at least on these grounds; von Neumann would certainly admit the single-particle and finite-particle cases as valid instances of his reversible ideal quantum gas thought experiment.

The historical component of Chua's argument is closely related to the validity component, in which Chua argues that the single-particle case does not constitute a valid thermodynamic system, and should therefore not be considered:

*"The Second Law, and hence phenomenological thermodynamics, should not be expected to hold true universally in small scale cases, and especially not in the single-particle case."* (19)

Since thermodynamics does not apply to the single-particle system, why should it matter that $S_{VN}$ fails to correspond to $S_{TD}$ in this case? The failure of the entropic arithmetic might even be expected.

*"H&S's reasoning is untenable, because they fail to respect the context of phenomenological thermodynamics by bringing it into a context where it is not expected to hold."* (19)

Chua is missing a crucial point here. As we saw in **Section 2.4**, von Neumann himself assumes the validity of the laws of thermodynamics in his thought experiments, even when considering the expansion of a single-molecule gas. Von Neumann follows Szilard's lead and assumes the validity of thermodynamic laws in order to study the statistical mechanical entropy (or classical entropy) $S_{CL}$. Thus, if his critique against H&S is to stand, Chua must also become a critic of von Neumann himself.

Furthermore, Chua concedes that the historical and validity components of his argument against H&S fail to hold if, instead of considering the correspondence between $S_{VN}$ and $S_{TD}$, H&S actually consider the correspondence of $S_{VN}$ and the subjectivist view of the Gibbs statistical mechanical entropy. He continues:

*"My above argument against the misapplication of phenomenological thermodynamics does not seem to apply here, since this argument is being made in*

*the context of statistical mechanics and its particle picture, with no commitment to phenomenological thermodynamics [...] Assuming the above picture is plausible, a failure of correspondence between $S_{VN}$ and the information entropy provides evidence against the correspondence of $S_{VN}$ and $S_{TD}$"* (20-22)

proving H&S right. Unfortunately, Chua fails to understand that von Neumann is indeed arguing from the statistical mechanical view himself.

### 4.3: Prunkl comes closest to understanding von Neumann

In our summary of Prunkl's argument against H&S, we began with a discussion of Landauer's Principle. Prunkl argues against the possibility of a unitary reset of the z-spin measurement device in H&S's **Stage 6**. Her argument, while thorough, is unclear on a critical point: if H&S are mistaken in assuming a unitary reset of the measurement apparatus, is von Neumann mistaken as well?

We have reached another point at which a close reading of von Neumann's original text is crucial. Indeed, in **Stage 6** of his cyclic gas transformation, von Neumann claims that the $P_{[\phi_1]}, P_{[\phi_2]}, \ldots$ gases can all be transformed back into a $P_{[\phi]}$ gas reversibly, which is equivalent to assuming a unitary reset of the measurement apparatus. As detailed in **Section 2.1**, von Neumann very carefully proves the reversibility of the reset transformation in **Stage 6**.

In light of this, we are forced to adopt a Landauer-inspired skepticism toward von Neumann's thought experiment as well. If we refuse to do this, then our only other option, apart from conceding a unitary reset to both H&S and von Neumann, is to somehow allow von Neumann the unitary reset but to deny it for H&S, recognizing that H&S's single-particle experiment simply cannot be taken as a faithful reconstruction of von Neumann's original work.

If Landauer's Principle is embraced, as Prunkl argues it must be, the debate at hand is forced to either (a) deny von Neumann's rigorous mathematical proof of a unitary reset of the measurement apparatus in his thought experiment or (b) acknowledge that it is centered around an altogether different thought experiment than the original, and leave von Neumann behind. Neither option is desirable for a debate on the nature of von Neumann's original work.

If we put this issue aside, granting both von Neumann and H&S their unitary measurement apparatus reset, we can address the most important part of Prunkl's paper: the Appendix. Within, Prunkl raises the critical point that a single-particle reconstruction of von Neumann's thought experiment, *based on his own original words,* must constitute a statistical ensemble, indeed an N-particle ensemble, and must be treated as such. This simple yet profound ideat calls H&S's entire project into question.

It is very disappointing that this insight, which arguably trumps any other critique of H&S's single-particle reconstruction so far, is not explored more deeply and developed into a coherent rebuttal against H&S in Prunkl's paper.

# 5. CONCLUSIONS

## 5.1: Why we should continue to study von Neumann's thought experiment

I have argued above that the debate over the question "does the von Neumann entropy correspond to the thermodynamic entropy?" is misguided. The misunderstandings present do not negate its rich significance in the philosophy of thermodynamics/statistical mechanics and quantum mechanics. Instead of abandoning the debate should continue to ask, "does von Neumann's thought experiment succeed?"

Before concluding, I would like to summarize some of the important questions raised by H&S, Prunkl and Chua in the course of their arguments either for or against the success of von Neumann's thought experiment. These questions showcase the power of von Neumann's work and demand further study.

(a) *The validity of Von Neumann's thermodynamic assumptions*. As argued in **Section 3.4**, Chua's argument against H&S's assumption of the validity of thermodynamic laws in the single-particle case can rightly be leveraged against von Neumann himself. We must ask whether or not von Neumann is justified in assuming the validity of thermodynamic laws in his thought experiment. If not, then the integrity of the thought experiment, like that of Szilard, is called into question. This serious problem forces us to question the role of idealizations and thought experiments, and must be explored further in order to defend, or debunk, von Neumann's original work.

(b) *The nature of quantum mixed states.* The third component of Chua's argument against H&S, what I termed the arithmetical component, centers around the nature and interpretation of quantum mixed states.

Recall that for H&S, the single-particle thought experiment fails for collapse interpretations of quantum mechanics. The location measurement in **Stage 5** decreases the information entropy and collapses the mixed state into a pure state, decreasing $S_{VN}$ so that the entropy changes don't add up. The thought experiment goes through for no-collapse interpretations, however, because the location measurement doesn't change $S_{VN}$.

According to Chua, H&S are wrong to assume $S_{VN}$ decreases in collapse interpretations following the location measurement in **Stage 5**. He explains that H&S hold an "ignorance interpretation" of quantum mixed states. The ignorance interpretation understands mixed states to represent a lack of knowledge about a system, i.e., a system represented by a mixed state is actually in a pure state, but the observer does not know which. This is why the location measurement reveals a pure state, decreasing $S_{VN}$.

The ignorance interpretation "confuses classical and quantum ignorance" and has no place in this thought experiment, because here the mixed state actually represents the relative frequency of different pure states in a mixture, where sub-ensembles are given respective weights based on their relative frequency. Under this correct interpretation, the location measurement does not cause $S_{VN}$ to decrease in **Stage 5**. In other words, H&S's entropic arithmetic is incorrect, and

*"it seems quite irrelevant whether we adopt a collapse or no-collapse interpretation, because the collapse mechanism applies to superposed pure states, not statistical mixtures."* (27)

Who has the right interpretation of quantum mixed states, Chua or H&S? The debate concerning the use of the ignorance interpretation of quantum mixed states is fundamental to quantum statistical mechanics. Based on von Neumann's definition of the statistical operator (as reviewed in **Section 2.1**), Chua's interpretation seems to match von Neumann's here, but there is still room for debate in this fundamental issue.

(c) *Landauer's principle*. H&S claim, as does von Neumann, that at the end of one cycle of the reversible gas transformation "the measuring device need also be returned to its initial ready state. One can do that unitarily." (162) As discussed above, Prunkl argues that a unitary reset of a measurement device is not possible, because a Landauer-type reset, at cost, is unavoidable (see **Section 3.5**). The only way to unitarily reset a measuring device is to record its final state beforehand, which would require a second measuring device, then a third, and so on ad infinitum. If this is true, H&S's entropic accounting is incorrect (and von Neumann's, in fact!) Much more work can be done to determine if Landauer's principle is valid and if it applies here. If it does von Neumann himself needs to be corrected.

(d) *The nature of the Einstein gas and its implications*. As argued both here and in Prunkl's paper, the statistical mechanical nature of the Einstein gas used in von Neumann's cyclic gas transformation entails that the single-particle system must be modeled as a multiple-particle ensemble. It appears that H&S are blind to this fact. How would their single-particle reconstruction look after taking this into account? Could it be redeemed from some or all of its flaws?

In this paper, we took a closer look at **Chapter V** of von Neumann's *Mathematical Foundations of Quantum Mechanics* in order to understand the true object of his famous thought experiment. After reviewing this thought experiment, we worked through two additional examples from his text: the expansion of a single-molecule gas, and macroscopic detection of photons on photographic plates. This closer look taught us that the current debate in the literature- between Hemmo & Shenker and Chua, in particular- misunderstands his goal. Von Neumann's goal was not to establish the correspondence between $S_{VN}$ and the phenomenological thermodynamic entropy, but to establish the correspondence between $S_{VN}$ and the Gibbs statistical mechanical entropy. While Prunkl hints at this correct understanding, she fails to follow through and utilize it in her argument against H&S.

The widespread misunderstanding in the literature surrounding the von Neumann entropy should not discourage us from continuing to ask, "does von Neumann's thought experiment succeed?" As we have discovered, this study of the fundamental intersection of statistical mechanics and thermodynamics furnishes us with multiple opportunities to answer lingering questions in the philosophy of physics.

# APPENDIX: Von Neumann's thought experiment in his own words

| Stage | Von Neumann's Description |
|---|---|
| Stage 0 | *Measurement of U-gas. This step is not explicitly described by von Neumann until the page after he concludes his thought experiment:* "…if $U$ is a state, $U = P_{[\phi]}$, then in the measurement of a quantity $\Re$ whose operator $R$ has the eigenfunctions $\phi_1, \phi_2, \ldots$, it goes over into the ensemble $U' = [\ldots]$ and if $U'$ is not a state, then an entropy increase has occurred (the entropy of $U$ was 0, that of $U'$ is >0, so that the process is irreversible." ($\Delta S = -\sum_{n=1}^{\infty} w_n Nk \ln w_n$) |
| Stage 1 | "…our U-gas is composed of a mixture of the $P_{[\phi_1]}, P_{[\phi_2]}, \ldots$ gases of $w_1 N, w_2 N, \ldots$ molecules respectively, all in the volume $V$." |
| Stage 2 | "We add an equally large rectangular box **K'** on to **K**, and replace the common wall by two walls lying next to each other. Let the one be fixed and semi-permeable—transparent for $\phi_1$, but opaque for $\phi_2, \phi_3, \ldots$; let the other wall be movable, but an ordinary, absolutely impenetrable wall. In addition, we insert another semi-permeable wall which is transparent for $\phi_2, \phi_3, \ldots$ but opaque for $\phi_1$." |
| Stage 3 | "We then push [the impenetrable wall] and [the $\phi_1$ impermeable wall], the distance between them being kept constant […] no work is done (against the gas pressure), and no heat development takes place." |
| Stage 4 | "Finally, we replace the walls [impenetrable wall and two semi-permeable walls] by a rigid, absolutely impenetrable wall [between K and K'…] in this way the boxes **K**, **K'** are restored. There is however, this change. All $\phi_1$ molecules are in **K'**, i.e., we have transferred all these from **K** into the same sized box **K'**, reversibly and without any work being done, without any evolution of heat of temperature change." ($\Delta S = \sum_{n=1}^{\infty} w_n Nk \ln w_n$) |
| Stage 5 | "We now compress these isothermally to the volumes $w_1 V, w_2 V, \ldots$ respectively, which requires investments of mechanical work $w_1 NkT \ln w_1, w_2 NkT \ln w_2, \ldots$ and the transfer of those amounts of energy (as heat) to a reservoir" |
| Stage 6 | "Finally, we transform the $P_{[\phi_1]}, P_{[\phi_2]}, \ldots$ gases all into a $P_{[\phi]}$ gas (reversibly, cf. above, $\phi$ an arbitrarily chosen state.) We have then only $P_{[\phi]}$ gases of $w_1 N, w_2 N, \ldots$ molecules respectively, in the volumes $w_1 V, w_2 V, \ldots$ Since all of these are identical and of equal density $N/V$, we can mix them, and this is also reversible. We then obtain a $P_{[\phi]}$ gas of $N$ molecules in the volume $V$ (since $\sum_{n=1}^{\infty} w_n = 1$). Consequently, we have carried out the desired reversible process." ($S = 0$) |